\documentclass[10pt]{article}
\usepackage[explicit]{titlesec}
\usepackage{hyphenat}
\usepackage{ragged2e}
\RaggedRight

\usepackage{garamondx}
\usepackage[T1]{fontenc}
\usepackage{amsmath, amsthm}
\usepackage{graphicx}

\usepackage{soul}
\setul{.6pt}{.4pt}

\usepackage{geometry}
\usepackage{fancyhdr}
\usepackage[labelfont={footnotesize,bf} , textfont=footnotesize]{caption}
\makeatletter
\renewcommand\tagform@[1]{\maketag@@@ {\ignorespaces {\footnotesize{\textbf{Equation}}} #1.\unskip \@@italiccorr }}
\makeatother
\titleformat{\section}
  {\normalfont}{\thesection}{1em}{\MakeUppercase{\textbf{#1}}}
\titlespacing\section{0pt}{0pt}{-10pt}
\titleformat{\subsection}
  {\normalfont}{\thesubsection}{1em}{\textit{#1}}
\titlespacing\subsection{0pt}{0pt}{-8pt}

\makeatletter
\newcommand\sixteen{\@setfontsize\sixteen{16pt}{6}}
\renewcommand{\maketitle}{\bgroup\setlength{\parindent}{0pt}
\begin{flushleft}
\vspace{-.375in}
\sixteen\bfseries \@title
\medskip
\end{flushleft}
\textit{\@author}
\egroup}
\makeatother

\usepackage[sort&compress]{natbib}
\makeatletter
\renewcommand\@biblabel[1]{\textbf{#1.}\hfill}
\makeatother

\bibpunct{}{}{,~}{s}{,}{,}
\usepackage{subcaption}
\usepackage{tabularx}
\usepackage{multirow}
\usepackage{cleveref}
\usepackage{bm}

\newcolumntype{b}{X}
\newcolumntype{s}{>{\hsize=.3\hsize}X}
\newcolumntype{m}{>{\hsize=.4\hsize}X}

\title{Classifying Lensed Gravitational Waves in the Geometrical Optics Limit with Machine Learning}

\author{
Amit Jit Singh*$^{,a}$, Ivan S.C. Li$^{b}$, Otto A. Hannuksela$^{a}$, Tjonnie G.F. Li$^{a}$, and Kyungmin Kim$^{a}$\\ \medskip \bigskip
$^{a}$Department of Physics, The Chinese University of Hong Kong, Shatin, New Territories, Hong Kong \\ 
$^{b}$Department of Physics, Imperial College London, London SW7 2AZ, United Kingdom\\ \medskip \bigskip
Students: singhamitjit@link.cuhk.edu.hk*, ivan.li16@imperial.ac.uk${}$ \\
Mentors: otto.hannuksela@link.cuhk.edu.hk,  tgfli@cuhk.edu.hk, kyungmin.kim@cuhk.edu.hk}

\begin{document}

% Makes the title and author information appear.
\vspace*{.01 in}
\maketitle
\vspace{.12 in}

% Abstracts are required.
\section*{abstract}
Gravitational waves are theorized to be gravitationally lensed when they propagate near massive objects. Such lensing effects cause potentially detectable repeated gravitational wave patterns in ground- and space-based gravitational wave detectors. These effects are difficult to discriminate when the lens is small and the repeated patterns superpose. Traditionally, matched filtering techniques are used to identify gravitational-wave signals, but we instead aim to utilize machine learning techniques to achieve this. In this work, we implement supervised machine learning classifiers (support vector machine, random forest, multi-layer perceptron) to discriminate such lensing patterns in gravitational wave data. We train classifiers with spectrograms of both lensed and unlensed waves using both point-mass and singular isothermal sphere lens models. As the result, classifiers return \textit{$F_{1}$} scores ranging from $0.852$ to $0.996$, with precisions from $0.917$ to $0.992$ and recalls ranging from $0.796$ to $1.000$ depending on the type of classifier and lensing model used. This supports the idea that machine learning classifiers are able to correctly determine lensed gravitational wave signals.
This also suggests that in the future, machine learning classifiers may be used as a possible alternative to identify lensed gravitational wave events and to allow us to study gravitational wave sources and massive astronomical objects through further analysis.

% Keywords are required.
\section*{keywords} 
Gravitational Waves; Gravitational Lensing; Geometrical Optics; Machine Learning; Classification; Support Vector Machine; Random Tree Forest; Multi-layer Perceptron

\vspace{.12 in}

% Start the main part of the manuscript here.
% Comment out section headings if inappropriate to your discipline.
% If you add additional section or subsection headings, use an asterisk * to avoid numbering. 

\section*{introduction}

A total of six gravitational wave (GW) signals have been detected by LIGO and Virgo detectors at the time of writing.
The source of these signals were believed to originate from merger events of binary systems, such as binary black holes (BBH)~\cite{GW150914:2016prl,GW151226:2016prl,GW170104:2017prl,GW170608:2017apjl,GW170814:2017prl} or binary neutron stars (BNS).~\cite{GW170817:2017prl}
With these recent detections, we now have the opportunity to study astronomical objects and events through GW observations in addition to the traditional methods of electromagnetic (EM) wave observations.

\begin{figure}[h!]
    \centering
    \includegraphics[width=\linewidth]{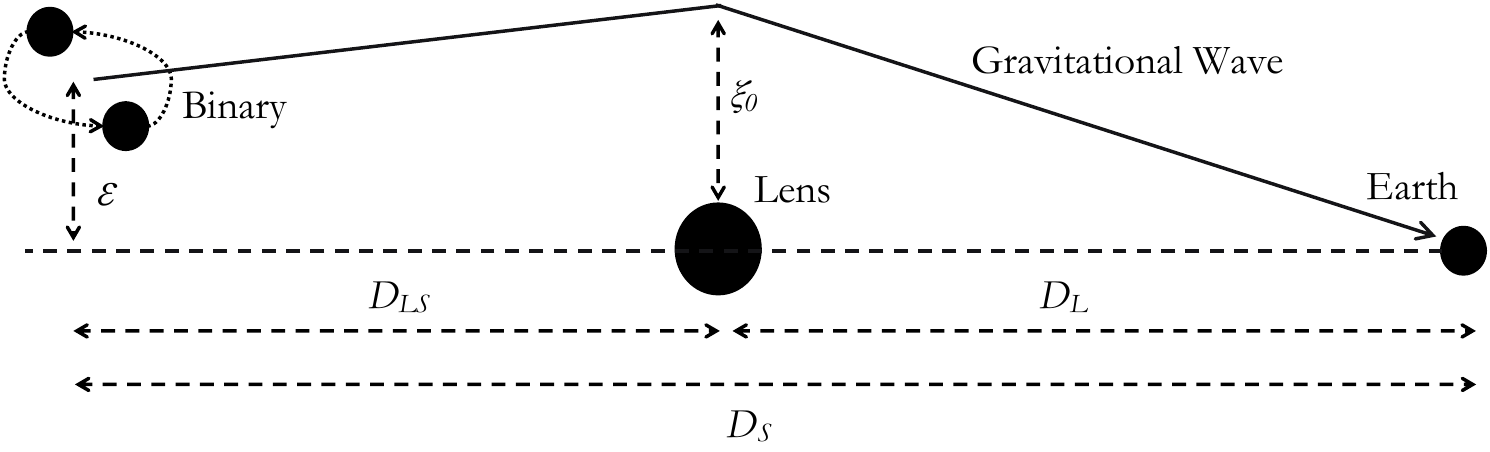}
    \caption{
    A simplified diagram of gravitational lensing of GWs.
    The binary system is the source of the gravitational waves.
    $\varepsilon$ is the distance of the binary from the line of sight extending from the 		lens and Earth.
    $D_{LS}$ is the distance from the source to the lens and $D_L$ is the distance from the 	lens to the Earth.
    $\xi_0$ is the Einstein radius of the lens.
    \smallskip}
	\label{lensingGWdiagram}
\end{figure}

When a GW signal passes by massive objects, the incoming wave behaves similarly to light in that the signal becomes gravitationally lensed (as shown in \textbf{Figure~\ref{lensingGWdiagram}}).
This changes the amplitude of the detected signal, and can cause multiple images from the same GW source to be detected at different times.~\cite{Treu:2014}
We consider two lensing models in this study: the point mass lens and the singular isothermal sphere (SIS).~\cite{Narayan:1996,takahashi:2003apj}
Our aim is to classify an incoming GW signal as lensed or unlensed through machine learning techniques.
By doing so, we may be able to better understand the properties behind the lens mass and GW source through further analysis of the GW signal.

Currently, the main method of identifying GW signals are through matched filtering techniques with pre-existing post-Newtonian GW waveform models.~\cite{GW150914:2016prl,GW151226:2016prl,GW170104:2017prl,GW170608:2017apjl,GW170814:2017prl,GW170817:2017prl,Owen:1998}
Some studies also suggest that matched filtering techniques can be used to obtain useful information such as the binary parameters of the GW source.~\cite{GW150914:2016prl,Cutler:1994,Cho:2015}
However, for this study, we focus on the ability of supervised machine learning classifiers (MLC) to determine if an incoming GW signal as a number of difficulties arise with the traditional match filtering techniques. With match filtering, we require accurate templates of the waveforms to compare with detected signals. However, the current lensing models are not exact and thus their templates may not match with actual lensed events, making the method relatively imprecise.
Furthermore, unlike unlensed GWs, we would require significantly more templates for lensed GW due to the different number of lens that exist and the increased number of physical parameters that come with lensing. On the other hand, machine learning does not require us to have a exact model and we do not have to concern ourselves with creating templates with various parameters due the probabilistic nature of the machine learning classifiers.
In particular, we choose to test three popular supervised MLC algorithms: support vector machine,~\cite{Cortes:1995ml} random forest,~\cite{Breiman:2001} and neural network.~\cite{Bishop:2017}

In this study, we generate spectrogram samples of lensed and unlensed GW signals under both the point mass lens and SIS models.
We use these spectrogram images as the training and testing data for these MLCs, and we analyze the performance of each MLC in the identification of lensed GWs by comparing the \textit{$F_{1}$} scores of each classifier.~\cite{Chinchor:1992:MEM:1072064.1072067}

\subsection*{Gravitational Lensing of Gravitational Wave Signals}
\label{sec:lens}

Models for lensing of GWs have been studied for decades.~\cite{Ohanian:1974, Bliokh:1975apss, Bontz:1981apss, Thorne:1983, Deguchi:1986apj, Schneider:1992, Nakamura:1999ptps, takahashi:2003apj}
From these studies, it is known that the lensing effect needs to be treated differently depending on the wavelength, $\lambda$, of a GW. We work in the geometrical optics approximation, where the wavelength of the incoming GW, $\lambda$, is much shorter than the Schwarzschild radius of the lens mass.
At longer wavelengths, the diffraction effect becomes dominant and the geometrical optics limit treatment of the GW lensing behaviour will become invalid.~\cite{takahashi:2003apj}
Throughout this study, we will be working in geometrized units (i.e. $G = c = 1$). 
We also make use of the thin lens approximation, where we assume that the GWs are scattered by a lens with its mass distributed across a two-dimensional plane perpendicular to the line of sight of the observer from the lens mass.~\cite{Narayan:1996}

In general terms, we can write the relation between the lensed GW, ${h_{\textrm{lens}}}(f)$, and the unlensed GW, $h(f)$ in frequency domain such that ${h_{\textrm{lens}}}(f) = F(f) h(f)$.
In this relation, the coefficient $F(f)$ is called the amplification factor and under the geometrical optics limit is defined as~\cite{takahashi:2003apj}
\begin{equation}
\label{eqn:ampfactor}
F(f) = \sum_j |\mu_{j}|^{1/2} e^{2\pi if \Delta t_{d,j} - i \pi n_j}, 
\end{equation}

where ${\mu_{j}}$ is the magnification factor of the j-th image of the lensed GWs, ${\Delta t_{d,j}}$ is the time delay of the j-th image, and ${n_j}$ is a discrete number which takes different values depending on the lens model.
For a given lens model, the magnifications and time-delays will differ.

In particular, the amplification factor is found by first using the lens surface mass density of a given lensing model to compute a deflection potential for the lens.
Finding the solution to an integral related to the deflection potential and various other source parameters would then yield an expression for $F(f)$.
The amplification factor is determined not only by the frequency of the GW, but also by the lens position and its mass distribution.

\subsection*{Point-Mass Lens Model}
For a point mass lens model, the lens mass is essentially defined as a two-dimensional Dirac-delta function on the thin lens plane.
Under this model, two images are detected by the observer, and \textbf{Equation~\ref{eqn:ampfactor}} reduces to~\cite{takahashi:2003apj}
\begin{equation}
\label{eqn:PMLampfactor}
F(f) = |\mu_{+}|^{1/2} - i |\mu_{-}|^{1/2} e^{2\pi if \Delta t_{d}}.
\end{equation}

By considering the red-shifted lens mass ${M_{Lz} = M_L (1 + z)}$, we can define the magnification factors of the two interfering GW images as ${\mu_{\pm} = \frac{1}{2}\pm\frac{(y^2+2)}{(2y\beta)}}$ and the time delay of the images as ${\Delta t_d = 4M_{Lz}\left[\frac{y\beta}{2}+\ln\left(\frac{\beta+y}{\beta-y}\right)\right]}$ where $\beta = \sqrt{y^2+4}$.
$y$ is a value which parameterizes the relative displacements of the source position, observer, and lens position. This is given by 
\begin{equation}\label{eqn:y_parameter}
y = \frac{\varepsilon D_L}{\xi_0 D_S} ,
\end{equation}

where $\varepsilon$ is the distance of the source from the line of sight, ${D_L}$ is the distance from the lens to the observer, ${D_S}$ is the distance from the source to the observer, and ${\xi_0}$ is the Einstein radius of the lens.

\subsection*{Singular Isothermal Sphere Model (SIS)}
Unlike the point mass lens, the singular isothermal sphere (SIS) model assumes that the lens mass is circularly and symmetrically distributed across the thin lens plane. 
It is generally used to model extended objects such as large stars or galaxy clusters. 
The SIS model is usually characterized by a velocity dispersion \emph{v}, which directly relates to the mass distribution of the model.~\cite{Narayan:1996,takahashi:2003apj} 
This consequently gives rise to different expressions for the magnification factors and time delay of the interfering GW signals.
It should be noted that under the SIS model, there is no second image detected by the observer if the lensing happens outside of the Einstein radius. 
Hence, the amplification factor for the SIS model is adjusted to become~\cite{takahashi:2003apj}
\begin{equation}\label{eqn:SISampfactor}
F(f) =
\begin{cases}
|\mu_{+}|^{1/2} - i |\mu_{-}|^{1/2} e^{2\pi if \Delta t_{d}} & \textrm{if $y<1$}, \\
|\mu_{+}|^{1/2} & \textrm{if $y\ge1$ ,}
\end{cases}
\end{equation}

where ${\mu_{\pm} = 1/y \pm 1}$ and ${\Delta t_d = 8M_{Lz}y}$.
In this case, $M_{Lz}$ is defined as the mass inside the Einstein radius of the lens mass.
The expression for the $y$ parameter in the SIS model is equivalent to that of the point mass lens model, given by \textbf{Equation~\ref{eqn:y_parameter}}.

\subsection*{Machine Learning Algorithms}

\label{sec:mlc}
We select three supervised classifiers which vary notably in their underlying algorithms for classification and prediction to find the most optimal method for analyzing spectrograms.
We choose to utilize the support vector classifier (SVC), the random forest classifier (RFC) and the multi-layer perceptron classifier (MLP). Below we give a general overview of how each algorithms works in order to highlight the significant difference in their method of classification which contributed to them being utilized.

Furthermore, we do not focus on the mathematical formulations of each classifier as they are not pertinent to the overall aim of this research. Readers interested in the mathematics of these algorithms are directed towards the references mentioned below.

\subsection*{Support Vector Classifier (SVC)}
SVC is based on the the support vector machine algorithm. Each item in our training set is plotted as a point in a \textit{n}-dimensional space where \textit{n} is the number of features representing an item and each feature is a coordinate of the item in the \textit{n}-dimensional space. The algorithm calculates a hyper-plane which divides the two classes of spectrograms and acts as a decision boundary. When we introduce the training set to the classifier, it predicts if a gravitational wave in a spectrogram is lensed based on where it is in the \textit{n}-dimensional space with respect to the hyper-plane.\cite{Geron:2017,Cortes:1995ml}

The two most important hyper-parameters that we consider are \textit{C} and \textit{$\gamma$}. \textit{$\gamma$} determines how the distance of each item in the training set influences the decision boundary. When \textit{$\gamma$} is large, the items closer to the boundary carry a larger weight and influence it heavily while the ones further away do not have much influence on its shape. On the other hand, for a small \textit{$\gamma$}, items nearer to the boundary are given less influence as the ones further away are also given importance. \textit{C} is responsible for determining the cost of a smooth decision boundary against the cost of misclassification of training points. If \textit{C} is large, the algorithm focuses on generating a decision boundary which classifies all items in the training set correctly. A small \textit{C} prioritizes the smoothness of the decision function and allows for a softer boundary such that there are some items in training set that cross and overlap the decision boundary into the other class. \cite{Cortes:1995ml}

\subsection*{Random Forest Classifier (RFC)}

The RFC is based on decision tree algorithms. A decision tree consists of a root node, which would be the training set, which splits into decision nodes based on discriminating features found in the data by the algorithm. These sub-nodes will continue to split until only the terminal node remains which would consist of a homogeneous piece of the original data and thus, cannot be split any further. When a new test item is introduced to the decision tree, it will classify the item according to which terminal node the data is matched to. In a RFC, multiple decision trees are grown. When making a prediction on a item in the test set, the most popular classification given by all the tree is used. This reduces the chance of error as multiple random decision tree are used rather than just one absolute one.\cite{Breiman:2001}

For this classifier, we consider two hyper-parameters. The first is the number of trees in the forest ($N_{t}$). More trees lead to better predictions but also utilizes more computational power. Thus, tuning it is required to avoid using unnecessary computational resources. The second is the minimum number of samples required to allow a node to split ($N_{\textrm{mss}}$). This is an important factor in controlling over-fitting as a small value could lead to the algorithm selecting highly specific features only occurring in a small set of one class and thus, do not generalize well over the whole class. A large value would then quite clearly lead to under-fitting.\cite{Breiman:2001}

\subsection*{Multi-layer Perceptron Classifier (MLP)}

The multi-layer perceptron classifier is a feed-forward neural network. The neural network consists of multi-layers as suggested by its name. The first layer is the input layer which consists of neurons matching the number of features of the input data. After the input layer, there are a number of hidden layers. In each layer, each neuron performs a linear weighted summation and then, a bias is added to avoid zero values. A non-linear activation function is applied to this summation and forwarded to the neurons in the next layer where this is repeated and continues until the outer layer is reached. At the outer layer, the output is given. However, the output often does not match the actual expected output. Thus, the backpropagation is applied to fix this. Backpropagation aims at minimizing the loss function using gradient descent. The loss function calculates the numerical difference between the actual class and the predicted class from the network. The loss is then minimized by adjusting the weights and bias using gradient descent which partially differentiates the loss function with respect to all the weights and bias. Each repetition of feed-forwarding and back-propagation is called an epoch. The MLP algorithm repeats this for a certain number of epochs until efficiency is achieved.\cite{Bishop:2017}

Although the MLP classifier has a number of hyper-parameters which can be optimized, we only consider the number of layers and the number of neurons in each layer ($N_{layer\_neurons}$) and the L2 penalty term $\alpha$. $\alpha$ is used to regularize over-fitting by adjusting the size of the weights. A larger $\alpha$ leads to smaller weights and creates a smoother decision boundary which reduced over-fitting. On the other hand, a larger $\alpha$ causes larger weights and a more complex decision boundary.\cite{Bishop:2017}
The rectified linear unit (relu) function is used as the activation function.

\section*{methods and procedures}

\subsection*{Gravitational Wave Model}

The unlensed waveform that we consider is from a binary inspiral source.
To simulate the lensed GW signal detected by the observer, we inject two GW strains $h(t)$ with a given amplification factor and time delay depending on the lensing parameters and the model used (point mass lens or SIS).
In particular, we generate a waveform computed up to $0.5$ post-Newtonian (PN) order.
This approximation should be sufficient for the purposes of this study as we are simply investigating whether the machine learning classifier is able to differentiate between a lensed and unlensed waveform.
For this reason, we also choose to omit the post-merger waveform from our model.
We use the following expression for the GW waveform \cite{Creighton:2011} 
\begin{equation}
h(t) = -8 \sqrt{\frac{\pi}{5}}\frac{\mu}{D_S}e^{-2i\varphi(t)}x(t) ,
\end{equation}

where $\mu$ is the reduced mass of the binary inspiral source.
To $0.5$ PN order, $x(t) = \varTheta(t)^{-1/4}/4$ and $\varphi(t) = -\varTheta(t)^{5/8}/\eta$ are the post-Newtonian parameter and orbital phase respectively.
$\varTheta(t) = \eta t/5M$ is a surrogate time variable, where $M$ is the total mass of the binary system and $\eta = \mu/M$ is the symmetric mass ratio.
\medskip

\begin{table*}[ht!]
\begin{center}
	\begin{tabularx}{1\textwidth}{s m m m m s s}
    \hline
    \hline
    \textbf{Parameter} & $\boldsymbol{M_L}$ & $\boldsymbol{D_L}$ & $\boldsymbol{D_{LS}}$ & $\boldsymbol{m_1, m_2}$ & $\boldsymbol{\varepsilon}$ & $\boldsymbol{z}$ \\
    \hline
    \textbf{Range} & $10-10^{7} M_\odot$ & $10-1000~Mpc$ & $10-1000~Mpc$ & $4-35 M_\odot$ & $0-0.5~pc$ & $0-2$\\
    \hline
    \hline
	\end{tabularx}
	\caption{
Ranges of randomized parameters of the GW waveforms.
$m_1$ and $m_2$ are the masses of the binary source, $D_{LS}$ is the distance between the source and lens mass, and $z$ is the redshift parameter.
The source and lens masses are sampled from a logarithmic distribution to reduce the bias towards more heavily lensed waveforms being generated.
    }
    \label{tab:GWwaveformparameters}
\end{center}
\end{table*}

\begin{figure}[t!]
  \centering
  \begin{subfigure}[b]{0.49\linewidth}
    \includegraphics[width=\linewidth]{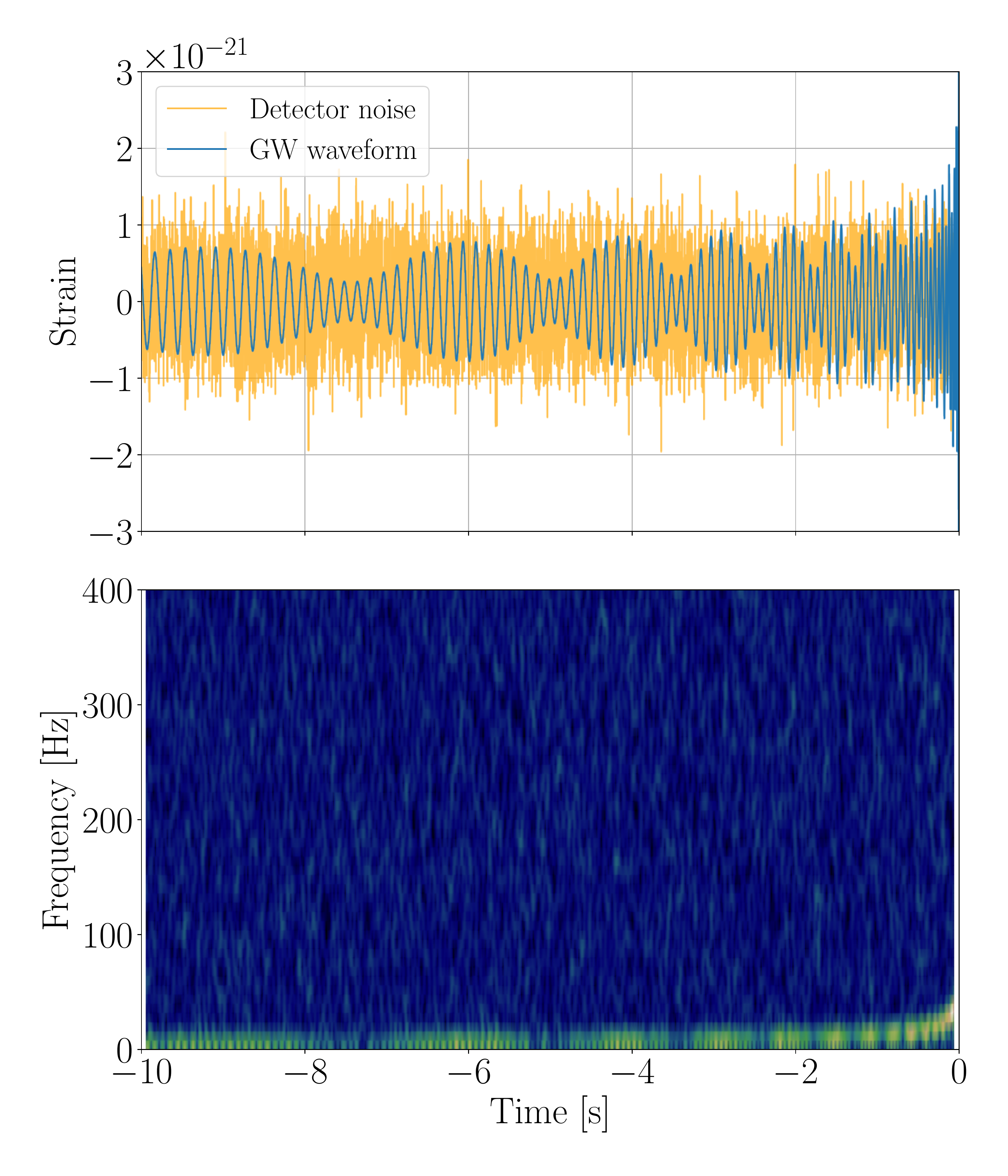}
    \caption{Lensed Gravitational Wave}
  \end{subfigure}
  \centering
  \begin{subfigure}[b]{0.49\linewidth}
    \includegraphics[width=\linewidth]{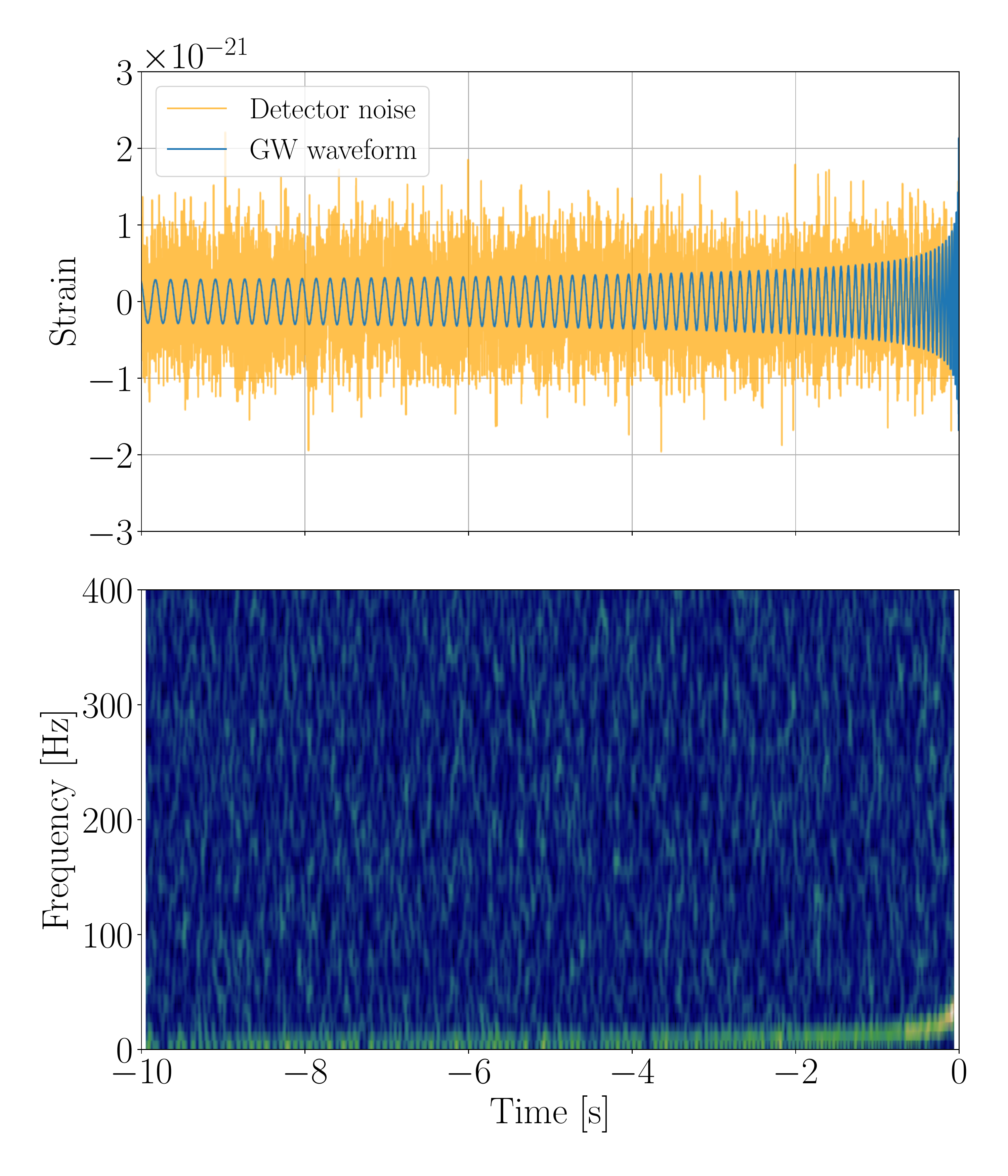}
    \caption{Unlensed Gravitational Wave}
  \end{subfigure}  
  \caption{
\emph{Top left:} A lensed GW waveform under the point mass lens model (with noise, colored in orange) generated from a binary inspiral source.The two interfering GWs that form the resultant lensed waveform are detected by the observer at different times, causing a beating effect to be seen. \emph{Bottom left:} A spectrogram of the incoming lensed signal is generated by performing a short-time Fourier transform on the signal. Coalescence of the binary system occurs at ${t=0}$, as seen from the increase in frequency of the GW over time. The same beating effect due to the lensing of the waveform can be observed. The spectrogram images are then used as the training data for the classifier. \emph{Top right:} An unlensed GW waveform with noise. The amplitude of the signal is smaller due to the lack of any lensing magnification. \emph{Bottom right:} A spectrogram of the unlensed signal. No beating effect is seen, and the smaller waveform leads to the noise being a more prominent feature in the spectrogram.
\medskip}
  \label{fig:GWspec}
\end{figure}

The resultant waveform is shown in \textbf{Figure~\ref{fig:GWspec}}. A beating effect is clearly evident due to the two interfering waveforms arriving at the observer with a time delay between them. In our model, the merger of the two component masses in the binary system occurs at ${t=0}$ , which can be seen by the peak in the GW waveform.
Unlensed waveforms are simply generated by injecting a single GW strain instead of two.
We randomize the parameters of the lensed and unlensed GW waveform within the ranges presented in \textbf{Table~\ref{tab:GWwaveformparameters}}.
We also implement Gaussian noise with an amplitude of the order $10^{-21}$ to test the ability of the classifier to identify a lensed waveform within a noisy background signal.

\subsection*{Signal-to-Noise Ratio (SNR)}
The signal-to-noise ratio (SNR) is a measure of the magnitude of a GW signal in relation to the background noise detected.
For a given GW waveform $h(f)$, we calculate the SNR using~\cite{Creighton:2011}
\begin{equation}\label{eqn:SNR}
\text{SNR} = \sqrt{4\int_{0}^{\infty} \frac{|h(f)|^2}{S_n(f)} df}
\end{equation}

where $S_n(f)$ is the power spectral density of the noise profile in the signal.
The SNR from the recent six GW detections ranged between $13$ to $32.4$.~\cite{GW150914:2016prl,GW151226:2016prl,GW170104:2017prl,GW170608:2017apjl,GW170814:2017prl,GW170817:2017prl}
In our study, we choose to limit the SNR of the generated data to be less than 80.
This is to ensure that the data we use for our MLCs are physically valid, and that the generated waveforms do not dominate the noise in the signal.

\subsection*{Spectrogram}
Since the GW waveform is generated in the time domain, we perform a short-time Fourier transform (STFT) on the data to extract the frequency information of the signal as a function of time.
We then create a spectrogram of the incoming signal with the Gaussian noise added, as shown in \textbf{Figure~\ref{fig:GWspec}}.
Depending on the lensing parameters used, the beating effect due to gravitational lensing (seen in \textbf{Figure~\ref{fig:GWspec}}) varies and is not always distinct even though the waveform is lensed. 
We prepare $2000$ spectrogram samples of lensed GWs ($1000$ each for the point mass and SIS lensing model) and $1000$ samples of unlensed GWs, which are used in the training and testing processes for the classifier.
Using \textbf{Equation~\ref{eqn:SNR}}, we find that our overall data has SNR values with a mean of $41$ and a standard deviation of $19$.

\subsection*{Optimization of hyper-parameters}
Before we train our classifiers, we perform a grid search with cross-validation to determine the optimal combination of hyper-parameters for each classifier. The result of the grid search are presented in \textbf{Table~\ref{hyperparams}}. %cite

\begin{table*}[h!]
	\begin{center}
		\begin{tabularx}{1.0\columnwidth}{X X X}
        	\hline
            \hline
        	\textbf{Classifiers} & \textbf{Hyper-parameters} & \textbf{Value}\\
            \hline
            \multirow{2}{*}{\textbf{SVC}} & $C$ & $1000$\\
            {} & $\gamma$ & $5\times10^{-7}$\\
            \hline
            \multirow{2}{*}{\textbf{RFC}} & $N_{t}$ & $1000$\\
            {} & $N_{\textrm{mss}}$ & $2$\\
            \hline
            \multirow{2}{*}{\textbf{MLP}} & $\alpha$ & $0.72$\\
            {}& $N_{layer\_neurons}$ & $1000$\\
            \hline
            \hline
		\end{tabularx}
        \caption{Optimal hyper-parameters for each classifier which are determined using a grid search with cross-validation. The hyper-parameters which gave the best score were chosen.}
        \label{hyperparams}
	\end{center}
\end{table*}

Applying the hyper-parameters stated in \textbf{Table~\ref{hyperparams}}, we train each classifier. Then, we use the trained classifier to predict the classes of the spectrograms in the test.

\subsection*{Evaluation of MLC performance}
We choose to assign $75\%$ of our spectrograms as the training set and retain the rest as the test set.
We perform two tests for each MLC, each time using different sets of spectrograms: first using the unlensed and point-mass-lensed spectrograms, then using the unlensed SIS-lensed spectrograms.
This results in two sets of data, one for each lensing model.
We analyze the results from the classifiers using the classification report and the receiver operating characteristic curve. 
The classification report provides several figure-of-merits (FOM): the precision, $\cal P$, recall, $\cal R$, and the \textit{$F_{1}$} score of the respective classifiers. These are defined as 
\begin{align}
{\cal P} &= \frac{\textrm{TP}}{\textrm{TP}+\textrm{FP}} \\
{\cal R} &= \frac{\textrm{TP}}{\textrm{TP}+\textrm{FN}} \\
F_1 &= \frac{2}{1/{\cal P}+ 1/{\cal R}}, \label{eqn:f1}
\end{align}

where TP is the number of true positives (i.e. correctly classified lensed waves), FP is the number of false positives (i.e. unlensed waves misclassified as lensed waves) and FN is the number of false negatives (i.e. lensed waves misclassified as unlensed). Precision is a measure of the accuracy of the positive predictions, whereas 
recall measures the sensitivity of the classifier in identifying lensed signals.
Precision and recall can be merged into a single metric called the \textit{$F_{1}$} score which is their harmonic mean, as shown in \textbf{Equation~\ref{eqn:f1}}.\cite{Geron:2017}

\begin{figure}[t!]
    \begin{subfigure}[b]{0.5\linewidth}
      \includegraphics[width=\linewidth]{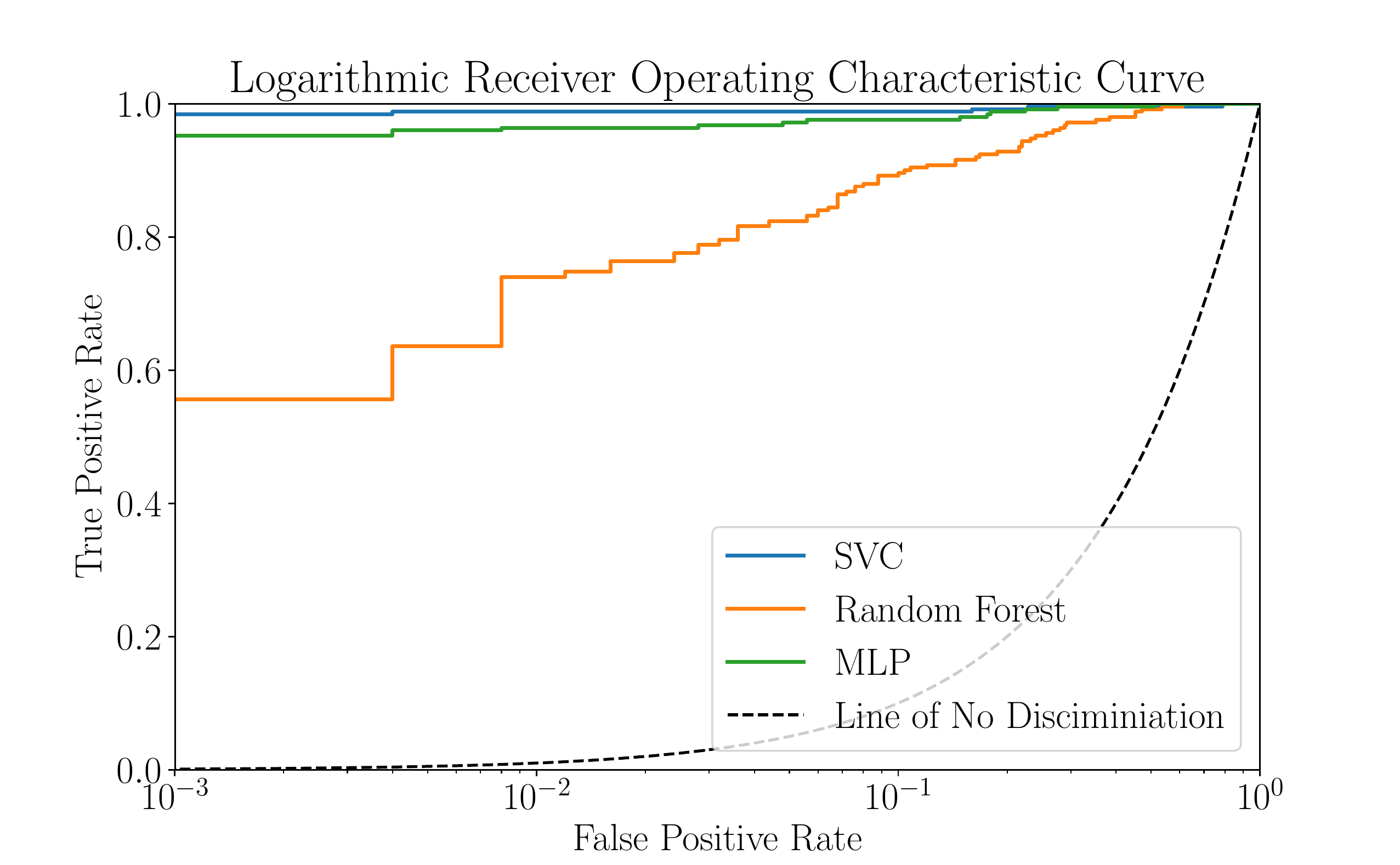}
      \caption{Point-Mass Lens Model}
    \end{subfigure}
    \begin{subfigure}[b]{0.5\linewidth}
      \includegraphics[width=\linewidth]{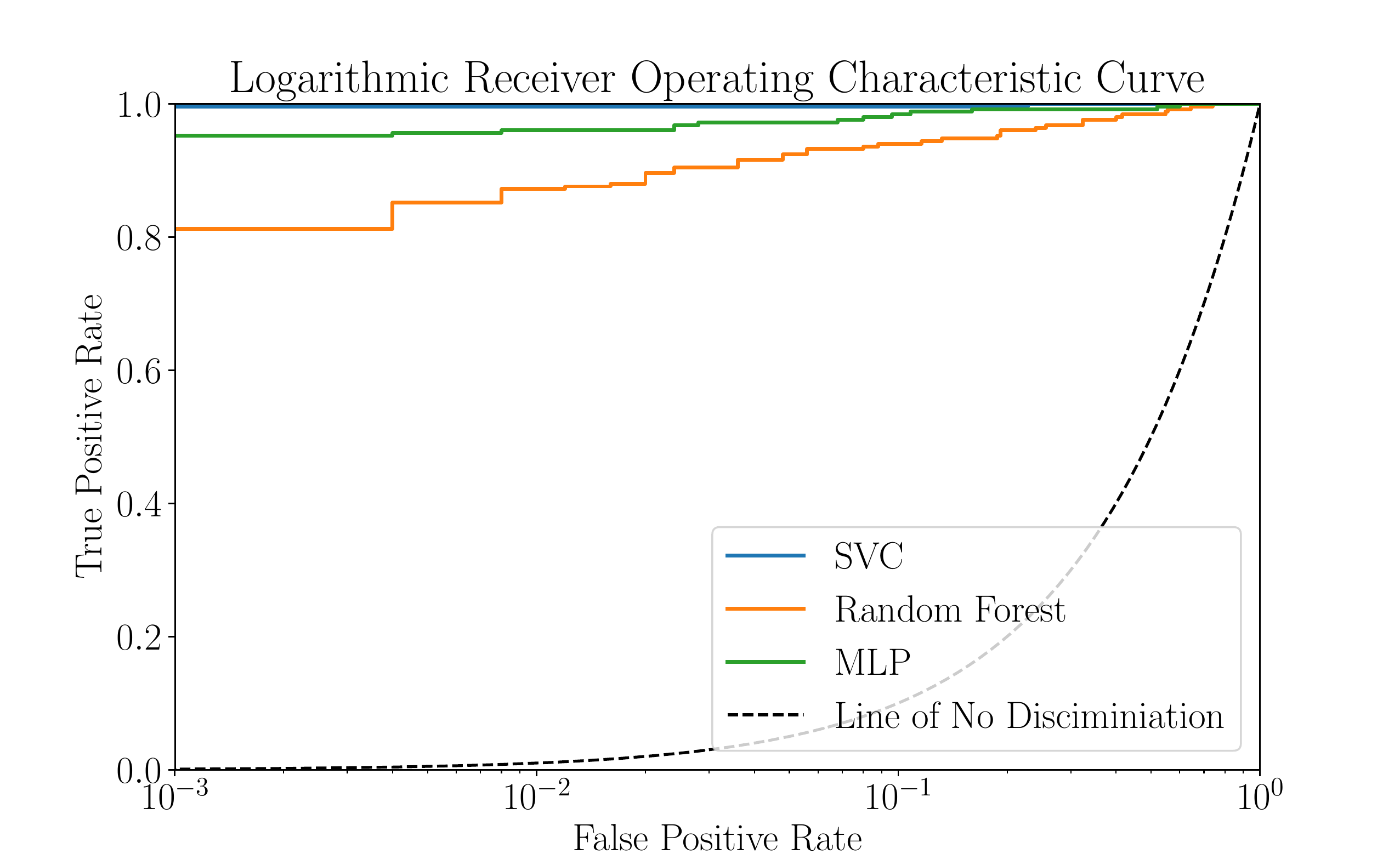}
      \caption{SIS Model}
    \end{subfigure}
  \caption{ROC curves of the three classifiers for both the point-mass lens and SIS lensing model.\bigskip}
  \label{fig:ROC}
\end{figure}

The receiver operating characteristic (ROC) curve is also used to find the most optimal classifier. The ROC curve plots the true positive rate against the false positive rate. The true positive rate is the same as recall while the false positive rate is the ratio of spectrograms were classified as unlensed but were lensed. The ROC curve provides us with the information whether the recall can be increased while keeping the false positive rate low. A theoretically perfect classification would allow us to increase the true positive rate to $1$ while maintaining the false positive rate at $0$. This also means the larger the area under the curve (AUC), the better the classifier is. By referring to \textbf{Figure \ref{fig:ROC}}, it would be clearer what this means.

\section*{results}
\begin{figure}[ht!]
  \centering
  \begin{subfigure}[b]{0.49\linewidth}
    \includegraphics[width=\linewidth]{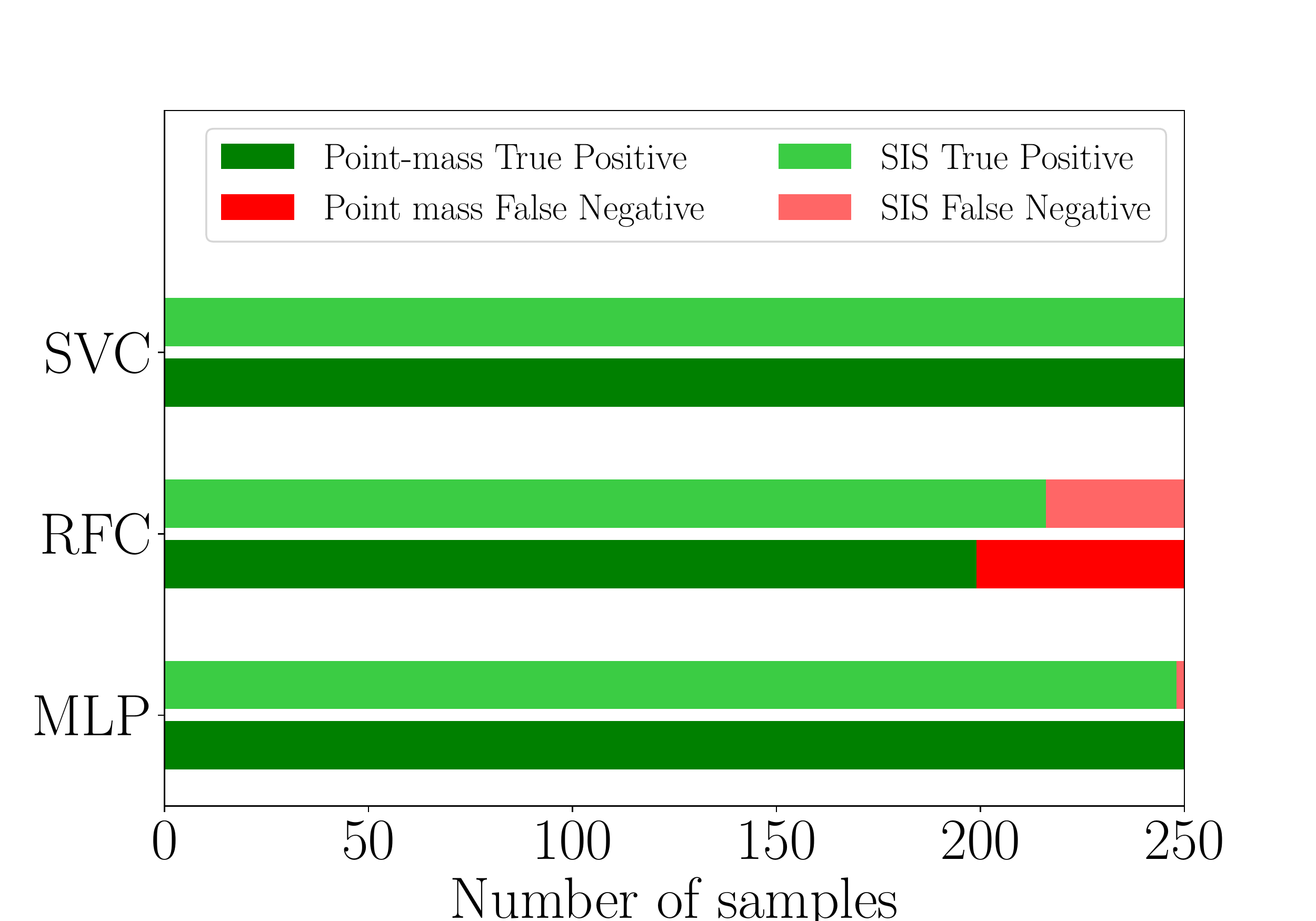}
    \caption{Lensed data}
  \end{subfigure}
  \begin{subfigure}[b]{0.49\linewidth}
    \includegraphics[width=\linewidth]{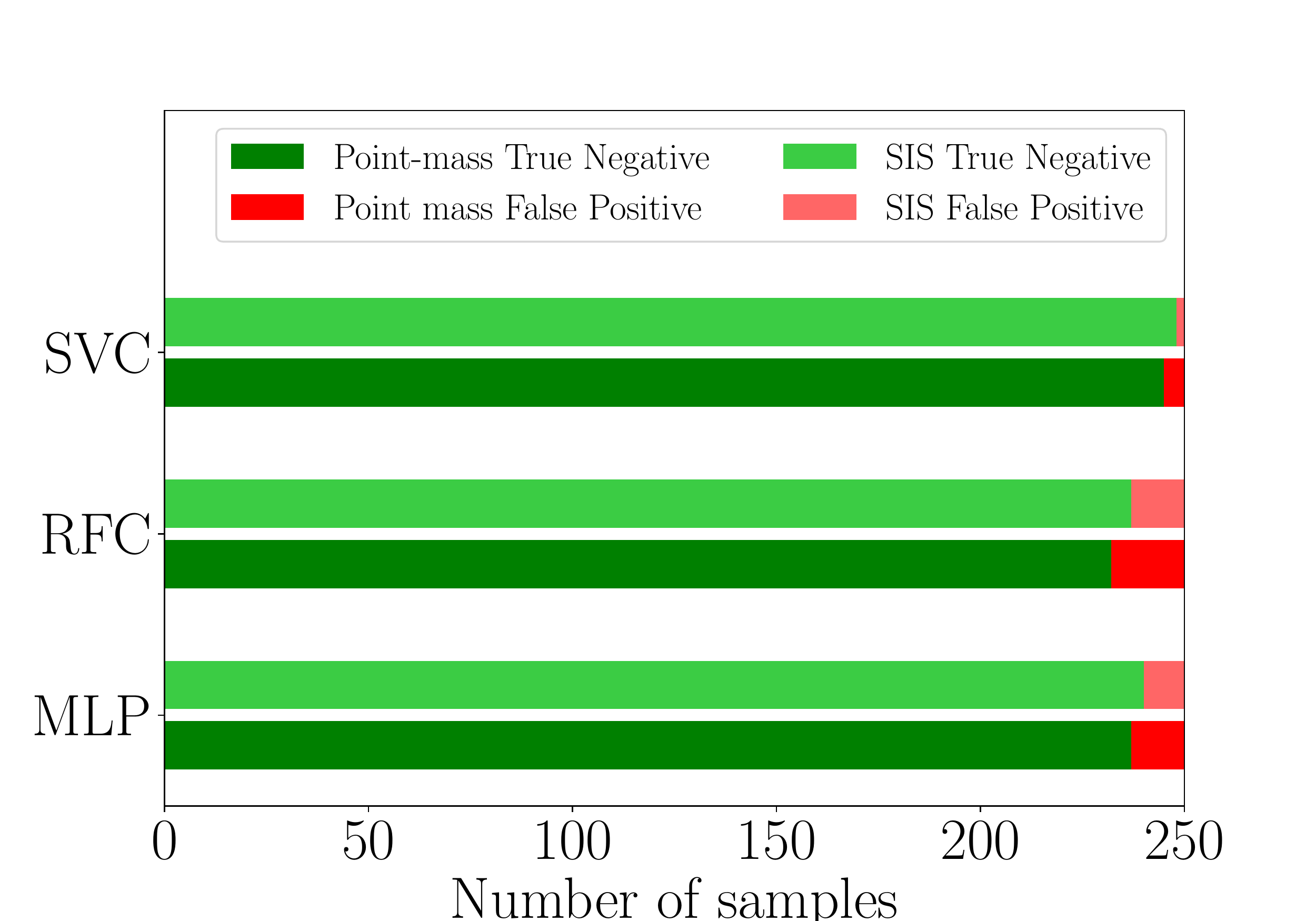}
    \caption{Unlensed data}
  \end{subfigure}
  \caption{Bar charts showing the performance of each MLC in classifying lensed and unlensed waveforms.
  The charts show the number of correctly classified samples against incorrectly classified samples for data from each lensing model.
  The RFC seems to be the most prone in misclassifying both lensed and unlensed waves, as seen from its high false positive and false negative rates for both the point mass and SIS model.
  The SVC and MLP classifier performed much better, with the SVC having a near overall perfect classification accuracy when classifying data from both lensing models.}
  \label{MLC_performance_chart}
\end{figure}

After the training and testing each classifier, the output data that we obtain includes a confusion matrix, a classification report, and an ROC curve for both the point-mass lens and SIS models.
The number of true positives, true negatives, false positives, and false negatives of each classification test are represented in a confusion matrix.
This information is then presented in a bar chart (\textbf{Figure~\ref{MLC_performance_chart}}) to allow us to compare the performances of each classifier.
As mentioned previously, we use $25\%$ of our data to test the MLCs, meaning that the testing data for each lensing model includes $250$ unlensed samples and $250$ lensed samples. 
In can be seen that the SVC had a false negative rate of zero when classifying data under both lensing models, as it was able to correctly identify all lensed samples in both lensing cases. The RFC seemed to be able to identify unlensed samples to a good degree of accuracy, but was generally weaker in classifying lensed signals as seen from its relatively high false negative and low true positive rates. The MLP classifier performed to a similar degree of accuracy to the SVC, but mistakenly classified two lensed spectrograms as unlensed under the SIS model. Using a larger data set and conducting a longer grid-search to identify the optimal hyper-parameters for each classifier may allow us to further investigate each of their strengths and shortcomings.

\iffalse
\begin{figure}[ht!]
  \centering
  \begin{subfigure}[b]{0.3\linewidth}
    \includegraphics[width=\linewidth]{SVC_cn_matrix_1000.pdf}
    \caption{Support Vector Classifier}
  \end{subfigure}
  \begin{subfigure}[b]{0.3\linewidth}
    \includegraphics[width=\linewidth]{RFC_cn_matrix_1000.pdf}
    \caption{Random Forest Classifier}
  \end{subfigure}  
  \begin{subfigure}[b]{0.358\linewidth}
    \includegraphics[width=\linewidth]{MLP_cn_matrix_1000.pdf}
    \caption{Multi-layer Perceptron Classifier}
  \end{subfigure}
  \caption{Confusion matrices for the three classifiers with Point-Mass Lens Model}
  \label{confusion_matrix_PM}
\end{figure}

\begin{figure}[ht!]
  \centering
  \begin{subfigure}[b]{0.30\linewidth}
    \includegraphics[width=\linewidth]{SVC_cn_matrix_SIS_1000.pdf}
    \caption{Support Vector Classifier}
  \end{subfigure}
  \begin{subfigure}[b]{0.30\linewidth}
    \includegraphics[width=\linewidth]{RFC_cn_matrix_SIS_1000.pdf}
    \caption{Random Forest Classifier}
  \end{subfigure}  
  \begin{subfigure}[b]{0.358\linewidth}
    \includegraphics[width=\linewidth]{MLP_cn_matrix_SIS_1000.pdf}
    \caption{Multi-layer Perceptron Classifier}
  \end{subfigure}
  \caption{Confusion matrices for the three classifiers with SIS Model}
  \label{confusion_matrix_SIS}
\end{figure}
\fi

The classification report gives the precision, recall, and $F_1$ scores of all three classifiers after fitting, as shown in \textbf{Table~\ref{tab:ClassificationReport}}. The MLC with the best performance seems to be the SVC, as it had $F_1$ scores of $0.990$ and $0.996$ when classifying lensed GW signals under the point-mass lens model and SIS model respectively. The RFC had lower $F_1$ scores as a result of its weaker recall ratios of $0.796$ and $0.864$. This was most likely due to its higher false negative rate meaning that the RFC was more likely to classify a lensed signal as unlensed. Again, the high $F_1$ scores of $0.975$ and $0.976$ for the MLP classifier show that it was able to identify both lensed and unlensed signals to a high degree of accuracy.

\begin{table}[t!]
	\begin{center}
		\begin{tabularx}{1.0\columnwidth}{X | X X X | X X X}
        	\hline \hline
        	\textbf{Lens Models} & \multicolumn{3}{>{\hsize=\dimexpr3\hsize+4\tabcolsep+2\arrayrulewidth\relax}X|}{\textbf{Point-mass Lens}} &  \multicolumn{3}{>{\hsize=\dimexpr3\hsize+4\tabcolsep+2\arrayrulewidth\relax}X}{\textbf{SIS}} 
            \tabularnewline \hline
        	\textbf{FOM} & \textbf{Precision} & \textbf{Recall} & \textbf{$F_{1}$ score} & \textbf{Precision} & \textbf{Recall} & \textbf{$F_{1}$ score}
            \tabularnewline \hline
            \textbf{SVC} &  $0.980$  &  $1.000$  &  $0.990$  &  $0.992$  &  $1.000$  &  $0.996$
            \tabularnewline \hline
            \textbf{RFC} &  $0.917$  &  $0.796$  &  $0.852$  &  $0.943$  &  $0.864$  &  $0.902$
            \tabularnewline \hline
            \textbf{MLP} &  $0.951$  &  $1.000$  &  $0.975$  &  $0.961$  &  $0.992$  &  $0.976$
            \tabularnewline \hline \hline
		\end{tabularx}
        \caption{
Classification report for all three MLCs, including all three figure-of-metrics (FOM). The SVC seemed to perform the best in identifying lensed signals under both lensing models, with $F_1$ scores of $0.990$ and $0.996$, while the MLP classifier performed slightly worse, with  $F_1$ scores of $0.951$ and $0.975$. The performance of the RFC suffered due to low recall ratios of $0.796$ and $0.864$.
        }
        \label{tab:ClassificationReport}
	\end{center}
\end{table}

\begin{table}[t!]
	\begin{center}
		\begin{tabularx}{1.0\columnwidth}{ X | X | X }
        	\hline \hline
            \multicolumn{3}{>{\hsize=\dimexpr3\hsize+4\tabcolsep+2\arrayrulewidth\relax}c|}{\textbf{Area under the curve (AUC)}}
            \tabularnewline \hline \hline
          	 & \textbf{Point-mass Lens} & \textbf{SIS} 
            \tabularnewline \hline
            \textbf{SVC} &  $0.995$  &  $0.999$
            \tabularnewline \hline
            \textbf{RFC} &  $0.962$  &  $0.976$
            \tabularnewline \hline
            \textbf{MLP} &  $0.993$  &  $0.993$
            \tabularnewline \hline \hline 
		\end{tabularx}
        \caption{The area under the ROC curve for the three classifiers in the two lensing models. The closer the area is to $1$, the better the classifier is for classifying the spectrograms.
        }
        \label{tab:AUC}
	\end{center}
\end{table}

The ROC curve for the SVC shows that we are able to increase the true positive rate (TPR) nearly to $1$ while maintaining to a false positive rate (FPR) close to $0$ for both lensing models. Similarly, the MLP classifier provides a relatively high TPR and low FTP. However, for the RFC,  reducing the FTP leads to a TPR below $0.6$ for the point-mass lens model and around $0.8$ for the SIS model. This indicates that even if we increase the recall of the classifier, the number of misclassifications would increase.

It can be seen that out of the three MLCs we tested, the SVC is the most accurate in terms of identifying a lensed GW signal.
It consistently has the highest precision, recall and \textit{$F_1$} score in both the point-mass lens and SIS model, with scores of over $0.980$ in all three metrics.
\textbf{Figure~\ref{MLC_performance_chart}} indicates that the SVC is more likely to misclassify an unlensed signal as lensed than misclassify a lensed signal as unlensed, although this result is statistically insignificant due to the limited data samples and the fact that the number of SVC misclassifications is still comparatively low.
On the other hand, the RFC has a higher chance of incorrectly predicting lensed GWs as unlensed.

The ROC curve for both models also indicates that SVC is the most appropriate classifier for our spectrograms for both model due to the high \textit{TPR} with a \textit{FPR} with the highest \textit{AUC} while the RFC performs the worst because even if we were to increase the recall of the classifier, the number of misclassifications would increase.
This means it has poor classification ability which is also reflected in the lowest \textit{AUC} in both models. This further implies that the SVC is the best classifier tested.

It should be noted that employing a grid-search and training the MLC takes significantly longer on the SVC, while it is the quickest on the RFC due to the nature of the algorithm itself and the fact that we could implement parallel processing when using the RFC.
Overall, all three MLCs were able to correctly differentiate between lensed and unlensed GW signals, albeit with varying degrees of accuracy.

\section*{discussion}
The use of machine learning classifiers is shown to be viable for identifying lensed GW signals, and can be seen as an alternative to the traditional matched filtering techniques.
However, there are many steps we could take to further improve our study and make our findings more rigorous.

One option is to use stricter assumptions in the generation of the GW signal.
As we are currently using $0.5$ PN order waveforms, we could instead consider using higher PN order waveforms provided by the PyCBC library to improve the accuracy of our GW waveforms.
Furthermore, an alternative to using pure Gaussian noise as the signal background is to use the noise power spectral densities provided by LIGO.
This would allow us to generate more realistic noise waveforms, as the characteristic noise profile of the LIGO detector has greater contributions at low and high frequencies, due to seismic noise and photon shot noise respectively.~\cite{AbramoviciLIGO,Abbott:2007LIGO}
Additionally, the SNR of our generated waveforms has a mean of $41$, which is arguably high when compared to the recent GW detections.
The binary source and lensing parameters may be fine-tuned in the future to reflect a more physically realistic GW source and lens mass, and ultimately produce GW signals with a lower SNR.
By reducing the the SNR values of the data to approximately $30$ or below, we will be able to test the limits of the MLCs' capability to identify lensed signals when the GW waveform is overwhelmed by noise. 

Due to the nature of grid-search processes and the complexity of MLC algorithms, computational and time limitations restricted our ability to use a larger dataset in this study.
It may be argued that the sample size we used for the training and testing processes for each classifier was too small to generate statistically significant results.
We are considering ways to implement a larger dataset of spectrograms with MLCs and reducing the size of the data such as utilizing principal component analysis and t-distributed stochastic neighbor embedding while maintaining a high accuracy of classification.
Another point of consideration is that we only completed a short grid-search for our hyperparameter selection process for each MLC. If we had performed a more rigorous and exhaustive grid-search, more optimal hyperparameters could be determined and we would be able to further investigate the extent of the MLCs' capabilities.

The implementation of alternative lensing models and MLCs to further test the validity of using machine learning to identify lensed waveforms may also be considered in the future.
A potential idea for future investigations may be to use machine learning to not only identify lensed signals, but to predict the properties of the GW source and lens mass associated with a given lensed GW signal through parameter estimation by employing regression algorithms in machine learning.

\section*{conclusions}
In summary, machine learning classifiers were able to identify lensed GW signals (with mean SNR = $41$) to a relatively high degree of accuracy.
This suggests that MLCs can be considered  as a viable alternative to matched filtering techniques in the search for lensing events.
Further investigations will be required to test the validity and reliability of the use of machine learning in classifying lensed GW signals, and to understand the limitations of this approach to a greater degree.
If subsequent research supports the validity of this method, a possible facet to explore in the future would be to use machine learning on a given GW signal to study and extract useful information regarding the associated GW source and lens mass.

\section*{acknowledgements}
This project and the students were supported by the 2018 Summer Research Program of the Department of Physics, The Chinese University of Hong Kong and was also partially supported by a grant from the Research Grants Council of the Hong Kong (Project No. CUHK 14310816 and CUHK 24304317) and the Direct Grant for Research from the Research Committee of the Chinese University of Hong Kong.

\bibliographystyle{unsrt}
\bibliography{gw,astro,method,lens}

\section*{about the student authors}
At the time of writing, Amit Jit Singh began his second year of undergraduate study at the Chinese University of Hong Kong where he is undertaking a BSc in Physics with plans to minor in Computer Science. He intends to attain a Ph.D. in Physics with a focus on data science.

Ivan S. C. Li is currently a third year undergraduate student at Imperial College London, where he is studying for an MSci in Physics. He undertook this research project at the Chinese University of Hong Kong as a summer intern.

\section*{press summary}
With the recent gravitational wave detections in the past few years, plenty of research has been focused on methods of examining and analyzing the waveforms to extract useful data. An interesting phenomenon that physicists study is gravitational lensing, which is where gravitational or electromagnetic wave signals can be lensed when they pass by massive astronomical objects. If we are able to identify gravitationally lensed signals, we may be able to learn more about the gravitational wave source and the massive object causing the lensing effect. Traditional methods of identifying lensed signals require highly precise and sophisticated lensing models, which we currently do not have. Our research aims to show that machine learning techniques pose as a possible alternative for performing this task. Through our study, we demonstrate that machine learning classifiers are able to differentiate between gravitationally lensed and unlensed gravitational-wave detections to a relatively high degree of accuracy, which suggests that machine learning techniques should be more frequently considered in gravitational wave analysis. However, further research is required to determine the extent at which machine learning is able to accurately extract useful information regarding the physical parameters of the gravitational wave source and lensing object.

\end{document}